\documentclass{actapoly}

\usepackage{aas_macros}
\usepackage[utf8]{inputenc}
\usepackage{natbib}

\newcommand*\printft[1]{\textsc{\lowercase{#1}}}

\begin{document}

\title[Reverse shock of GRB200131A]
{Early steep optical decay linked to reverse shock for GRB200131A}

\correspondingauthor[M. Jelínek]{Martin Jelínek}{asu}{mates@asu.cas.cz}
\author[F. Novotný]{Filip Novotný}{ipa,asu} 
\author[S. Klose]{Sylvio Klose}{tls} 
\author[B. Stecklum]{Bringfried Stecklum}{tls} 
\author[A. Maleňáková]{Alžběta Maleňáková}{mff} 
\author[J. Štrobl]{Jan Štrobl}{asu} 

\institution{asu}{Astronomical Institute of the Czech Academy of Sciences (ASU CAS), 25165 Ondrejov, CZ}
\institution{ipa}{Institute for Physics and Astronomy, University of Potsdam,Karl-Liebknecht-Str. 24/25, 14476 Potsdam, DE}
\institution{tls}{TLS Tautenburg, full postal address (street, postal code + city, country)}
\institution{mff}{Faculty of Mathematics and Physics, Charles University, full postal address (street, postal code + city, country)}

\begin{abstract}

We observed an optical afterglow of GRB 200131A obtaining the first photometric point 63\,s after the satellite trigger. 
This early observation shows a steep decay, suggesting either internal engine activity or a reverse shock.
By fitting this data set, we show that the early data fit well as a reverse shock component of the GRB afterglow modeled as a thin shell expanding into a constant density interstellar matter. 
The fitting also shows a good agreement with a catalogued Milky Way galactic extinction and leaves only little space for further extinction in the host galaxy. 
By judging several factors we conclude that the most likely redshift of this GRB is $0.9\pm0.1$. 


\end{abstract}


\maketitle

\section{Introduction}

The reverse shock (RS) is a short-lived, yet highly significant feature of gamma-ray bursts (GRBs \citep{bigreview} \citep{reverse}) within the context of the relativistic fireball model. It arises when the relativistic ejecta from the burst collide with the surrounding medium, creating a shock wave that propagates back into the ejecta, while a forward shock (FS) propagates into the external medium. The reverse shock is responsible for the production of prompt optical and radio emission, and can provide valuable insights into the physical conditions of the GRB outflow and its interaction with the environment. The relative strengths of the reverse and forward shocks depend on the properties of the ejecta and the surrounding medium, as described by the relativistic fireball model. However, the detection and characterization of reverse shocks have proven to be challenging due to their transient nature and the complex interplay of various emission processes.

In recent years, advancements in observational facilities, namely robotic telescopes with their very quick reactions to GRB alerts, allow for better data and therefore we may readily test the theoretical predictions for early GRB afterglow emission. 

\paragraph{GRB 200130A} was a long gamma-ray burst (GRB) detected by instrument BAT onboard {\it Swift} satellite on January 31 2020 at 22:41:16 UT \citep{gcnswift} in the southwest of Cassiopeia constellation, about 15$^\circ$ NE from M31. 
Simultaneous detection was made by instrument Konus onboard {\it Wind} satellite \citep{gcnkonus}. 
Soon, an optical counterpart was discovered by UVOT \citep{gcnuvot}. The alert was followed-up by a wide range of telescopes including RATIR \citep{gcnratir,gcnratir2}, LCO \citep{gcnlco} or VIRT \citep{gcnvirt}.

Our Small Binocular Telescope (SBT) in Ondřejov \citep{stroblsbt} promptly reacted to the received alert, slewed to the GRB coordinates and at 22:42:13.36\,UT (i.e. 57.4\,s after trigger) started obtaining a predefined imaging sequence.
We continued to follow up the source with the SBT in a clear filter until 30,min after the GRB. In total, we obtained 39 unfiltered exposures in each of the primary cameras C1 and C2 and 48 exposures with the auxiliary camera C3, with exposures varying between 10, 30 and 60\,s over the course of the observations. These frames were combined as necessary to provide a final set of 10 photometric points from the merging of the two primary cameras and 4 from the auxiliary camera.

Later, we obtained two sets of exposures
using the TLS Schmidt telescope and the TAUKAM prime focus camera \citep{taukam}.
Six full-frame images with an exposure time of 180\,s each were taken at each epoch. 
A wide V-band filter (VB) was used with a transmission curve
similar the Gaia GBP filter \citep{gaiaphot}.
Preliminary photometry and information was reported in the GCN \citep{gcntls}. 


\begin{table}[t]
    \centering
    \begin{tabular}{lccc}
    \toprule
dT [s] & T$_{exp} [s]$ & brightness [mag] & filter \\
\midrule
\multicolumn{4}{c}{\bf SBT camera C1 and C2} \\
63.1&    10& 14.155$\pm$0.038& cr\\
71.5& 30& 14.362$\pm$0.025& cr\\
120.1&    30& 15.256$\pm$0.040& cr\\
160.7&    30& 15.565$\pm$0.054& cr\\
201.3&    30& 15.956$\pm$0.092& cr\\
241.8& 30& 16.096$\pm$0.122& cr\\
282.4&    30& 16.226$\pm$0.264& cr\\
343.7&    71& 16.365$\pm$0.264& cr\\
547.0&    330&    16.845$\pm$0.284& cr\\
1338.9& 1201&   17.557$\pm$0.221& cr\\
\multicolumn{4}{c}{\bf SBT camera C3} \\
62.2&    10& 14.749$\pm$0.54& cr\\
83.8&    30& 14.693$\pm$0.17& cr\\
125.6&    63& 15.494$\pm$0.26& cr\\
208.1&    194&    15.894$\pm$0.42& cr\\
\midrule
\multicolumn{4}{c}{\bf TLS Tautenburg}\\
1860&	6$\times$180 &	18.00$\pm$0.10&	VB \\
79232&	6$\times$180 &	21.85$\pm$0.30&	VB \\
\midrule
\multicolumn{4}{c}{\bf RATIR \citep{gcnratir,gcnratir2}}\\
23580&  2952&   19.97$\pm$0.04&   $r'$\\
19260&  2952&   19.75$\pm$0.02&   $i'$\\
102870& 5346&   21.39$\pm$0.11&   $r'$\\
102870& 5346&   21.39$\pm$0.07&   $i'$\\
187146& 1764&   22.26$\pm$0.29&   $r'$\\
187146& 1764&   22.45$\pm$0.35&   $i'$\\
\midrule
\multicolumn{4}{c}{\bf LCO \citep{gcnlco}}\\
11610&  450&    19.75$\pm$0.13&   $R$\\
\midrule
\multicolumn{4}{c}{\bf VIRT \citep{gcnvirt}}\\
5760&   1&  18.8$\pm$0.2& $R$\\
\midrule
\multicolumn{4}{c}{\bf UVOT \citep{gcnuvot}}\\
152.5&	149&	15.47$\pm$0.02&	White\\
5203&	197&	19.53$\pm$0.16&	$b$\\
306.5&	32&	16.09$\pm$0.08&	$u$\\
4792&	197&	18.48$\pm$0.18&	$w1$\\
5305&	393&	18.98$\pm$0.20&	$m1$\\
4895&	393&	19.06$\pm$0.19&	$w2$\\
\midrule
\multicolumn{4}{c}{\bf CDK700 \citep{gcncdk}}\\
66297&	5040&	21.4$\pm$0.3&	$R$ \\
\midrule
\multicolumn{4}{c}{\bf ISON \citep{gcnison}}\\
3761&	3600&	18.35$\pm$0.25&	Clear \\


\bottomrule
    \end{tabular}
    \caption{Collection of optical photometric data for GRB 200131A.}
    \label{tab:phot}
\end{table}
\section{Observations and Data Reduction}

\subsection{Optical observations}

\paragraph{Astrometry}

We used the second set of Tautenburg images to measure the position of the detected optical afterglow. 
Our best-fit astrometric position is
\begin{center} 00:12:22.60 +51:07:00.2 $\pm$ 0.2$"$ \quad (J2000)\end{center} 
(3.094177 +51.116724). 
The astrometric solution was performed using 180 stars from the 
Gaia-DR2 catalogue \citep{gaia-dr2} 
and includes both position measurement and statistical error. 

\paragraph{Photometry} The calibration of the unfiltered photometric data is challenging. 
First, the zeropoint needs to be related to a standard system, e.g. SDSS AB magnitudes. 
This can be done by fitting the brightness of the detected objects as a function of catalogued stars.
We perform this task with our PYRT package \citep{jelinekpyrt}.
Here we expressed our unfiltered data as a function of $g'$, $r'$ and $i'$ entries of the Atlas \citep{atlas} photometric catalogue.
The brightness of the object, however, corresponds to a new photometric system defined by the width of its passband. 
To compare the afterglow brightness between this system and standard filters, we need to homogenize the result based on the known or assumed colour of the optical afterglow.
In our case, we could calculate the colour indices $(g'-r') = 0.36$\,mag and $(r'-i') = 0.25$\,mag from the later photometry, and assuming a constant spectrum, we derive the relative shift of the afterglow brightness ($r'_\mathrm{OA} - cr'_\mathrm{raw,OA} = +0.02$\,mag). 
The Tautenburg data were treated in a similar way.
%
The unfiltered measurements provided in Table\,\ref{tab:phot} as filters $cr$ and $VB$ include this correction, and can be directly compared to the $r'$-band, while keeping in mind the assumption of a constant spectrum. 

\paragraph{Image combination}

The photometric data combination process involves several steps to ensure reliable results while maintaining temporal resolution. Individual frames were combined using the Montage software package \citep{montage,montagesw}, with the specific combinations determined by an iterative process that optimizes signal-to-noise ratio (SNR) while preserving the temporal evolution of the afterglow. The astrometric solutions necessary for image resampling and the photometric solutions for weight determination were computed using the PYRT package \citep{jelinekpyrt}.

For each temporal bin, images were weighted based on their individual photometric quality, with weights derived from the statistical uncertainty of the photometric fit. The final temporal binning was selected to maintain SNR above a threshold that ensures reliable photometry while providing sufficient sampling of the afterglow evolution. Multiple combination trials with different temporal groupings were performed to verify that the final light curve accurately represents the afterglow behavior and is not biased by the specific choice of temporal bins.

\paragraph{Afterglow fitting}

The afterglow model fitting was performed using gnuplot \citep{gnuplot}, employing $\chi^2$ minimization. The fitting was conducted in magnitude space, which, while introducing some bias for the faintest detections, provides better numerical stability across the large dynamic range of the afterglow brightness evolution. The parameter uncertainties were determined using the standard error estimation from the inverted Jacobian matrix. Given the complexity of the model and the number of free parameters, careful human-guided selection of initial parameter estimates was crucial for achieving convergence to physically meaningful solutions. This approach, while more time-consuming than fully automated fitting, allows for better control of the parameter space exploration and helps avoid local minima that might not represent physically realistic solutions.

\begin{figure}
    \centering
    \includegraphics[width=0.75\linewidth]{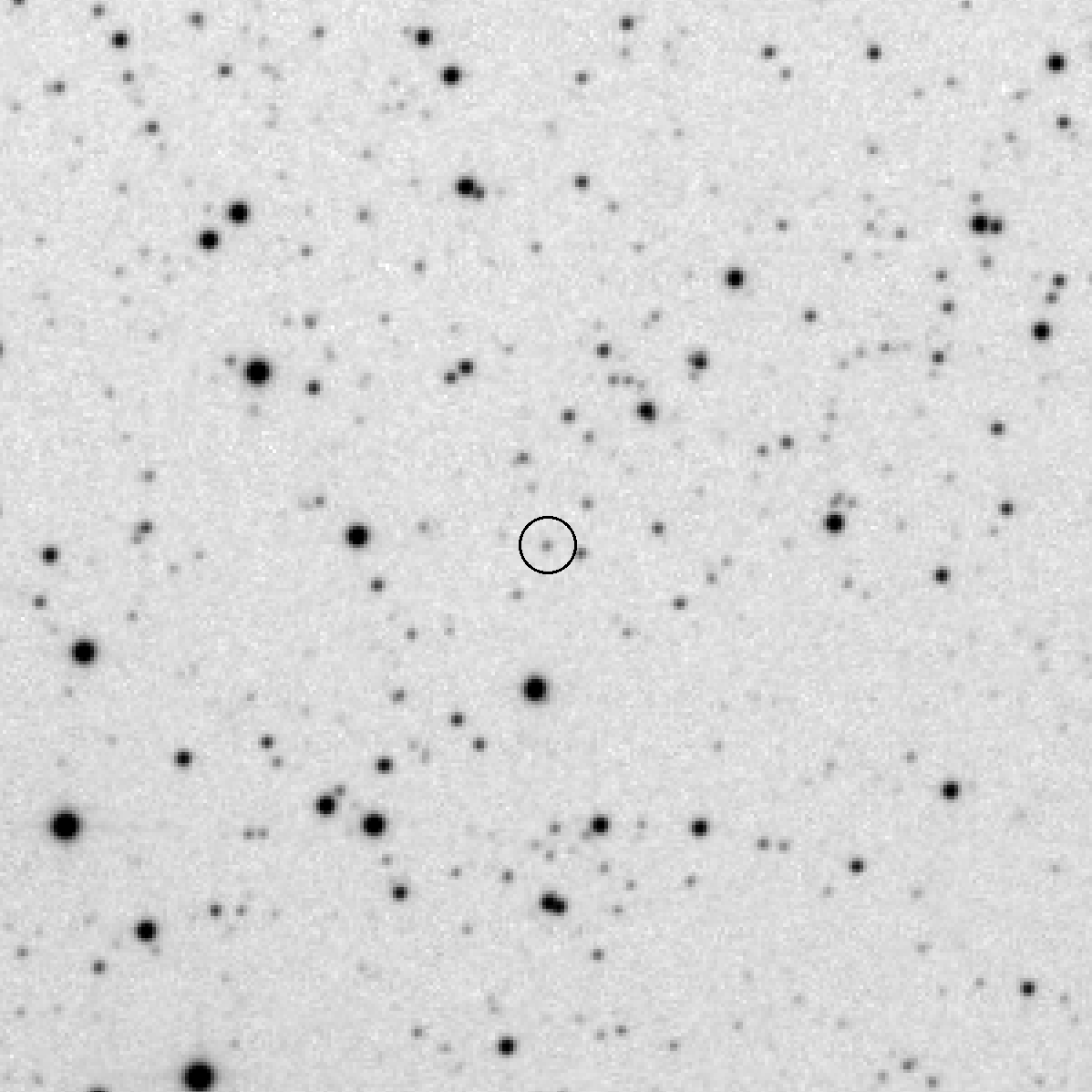}
    \includegraphics[width=0.75\linewidth]{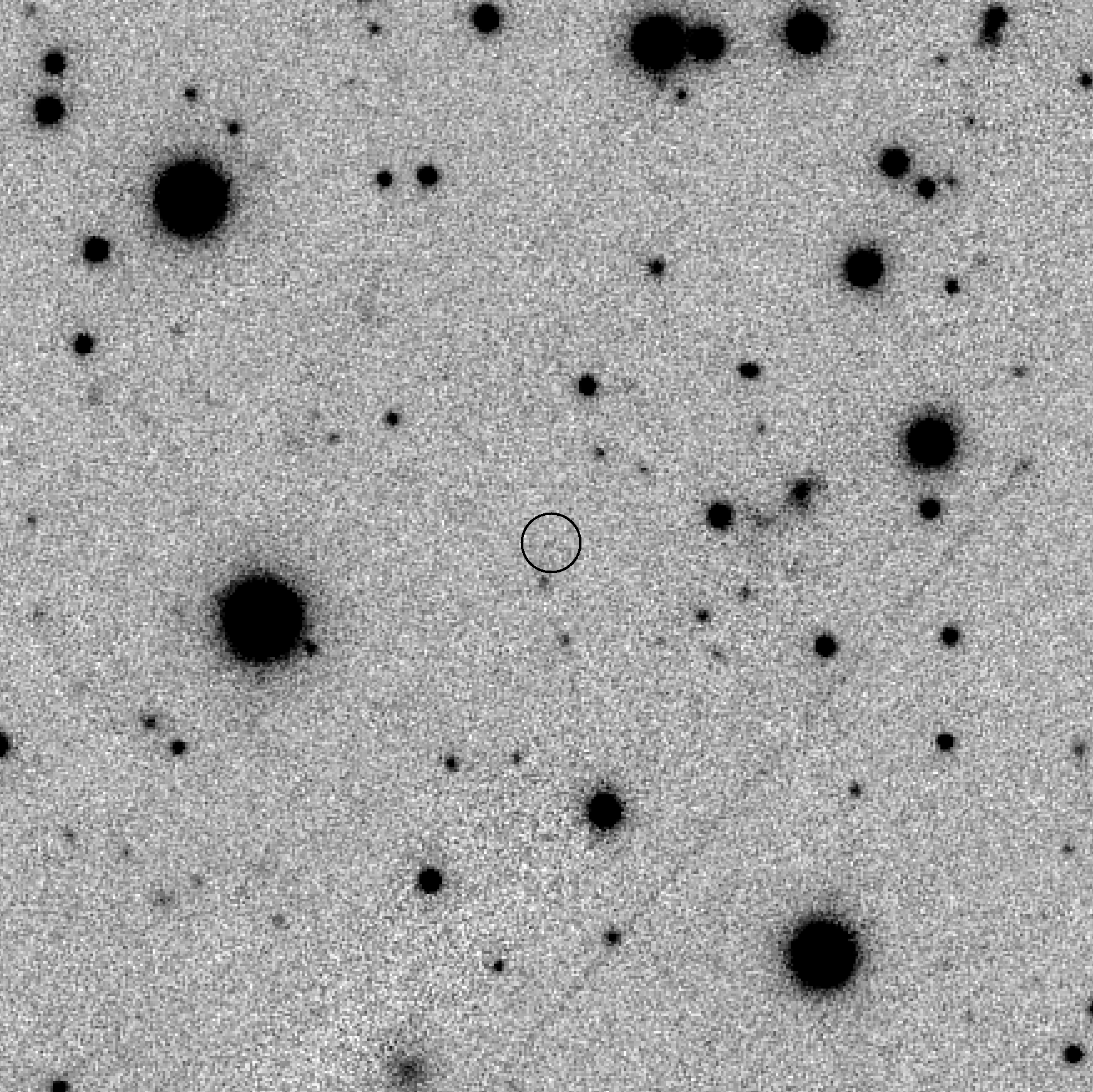}
    \caption{Top: Discovery image from the Ond\u{r}ejov SBT telescope, marking the optical afterglow of GRB\,200131A. The image covers {10}\,arcmin per side.
    Bottom: $ r' + i' + z' + y'$ image from the PanSTARRS archive, marking the location of the optical afterglow of GRB\,200131A. The underlying object is clearly visible. This image has a dimension of {$1.5\times1.5$\,arcmin$^2$}. }
    \label{fig:image}
\end{figure}

\begin{figure}
    \centering
    \includegraphics{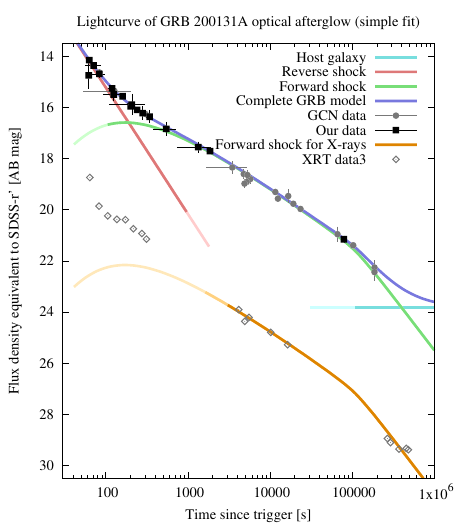}
    
    \caption{Light curve of GRB200131A fitted with a superposition (green) of two components - early reverse shock (red) and later forward shock (blue).
    \label{fig:lc}}
\end{figure}

\begin{figure}
    \centering
    \includegraphics{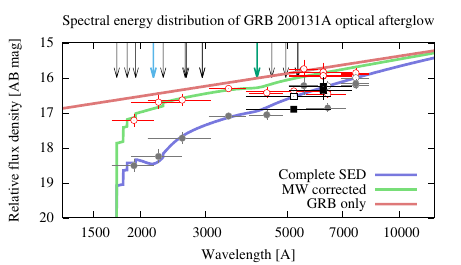}
\caption{Spectral energy distribution of GRB200131A, as fitted with the afterglow spectral slope derived from temporal decay, MW extinction, and host extinction (SMC type) with $z$ fixed to 0.9. Black points show observed photometric measurements, while red points show the same data corrected for Milky Way extinction. Black arrows from top indicate positions of typical intergalactic absorption features (Lyman series, Mg II, C IV), blue arrow marks Ly-$\alpha$, and green arrow indicates the characteristic Milky Way UV absorption feature. The model fit is shown considering both the intrinsic spectral slope and various extinction components.
\label{fig:sed}}
\end{figure}

\begin{table}[t]
    \centering
    \begin{tabular}{cc}
     \hline   
     parameter                         & fit \\ \hline
$p$ & 2.295          $\pm$ 0.017      \\
$A_\mathrm{v,host}$ & 0.084        $\pm$ 0.051\,mag       \\
$M_\mathrm{rs}$ & 17.07          $\pm$ 0.08\,mag      \\
$M_\mathrm{fs}$ & 14.00           $\pm$ 0.33\,mag       \\
$T_\mathrm{fs}$ & 33          $\pm$ 22\,s        \\
$G_\mathrm{fs}$ & 1.01         $\pm$ 0.20       \\
$T_\mathrm{jb}$ & 112155           $\pm$ 17600\,s    \\
$G_\mathrm{jb}$ & 0.16         $\pm$ 0.10       \\
$B_\mathrm{R}$ & 0.63        $\pm$ 0.10\,mag       \\
$B_\mathrm{TLS}$ & 0.366         $\pm$ 0.068\,mag      \\
$B_\mathrm{i'}$ & 0.211         $\pm$ 0.040\,mag      \\
\hline 
NDF &                        31 \\
WSSR/ndf   & 0.704 \\
\hline
    \end{tabular}
    \caption{\label{tab:fit1} Afterglow fitting parameters required by our model. Redshift is fixed to $z=0.9$. $M$ are scalings for reverse and forward shocks, $T$ are times of hydrodynamic and jet break, respectively. $G$ are smoothnesses of these breaks. $B$ values are necessary photometric shifts for parts of the data set.}
    \label{tab:fit}
\end{table}

\subsection{Host galaxy}


We searched the archival data of Pan-STARRS \citep{panstarrs} for detection of underlying emission at the location of the optical afterglow.
Pan-STARRS provides five channels in filters $grizy$. 
The images were taken in 2014, so no contamination from the afterglow should be expected. 
The single channels provide a threshold hint of a possible detection, after coadding all filters except for $g$ we obtain a 4-$\sigma$ detection with brightness $22.99\pm0.27$ mag (AB). 
The per-filer values measured from PanStarrs are $g>23.5$, $r = 23.6\pm0.4$ and $i=23.7 \pm 0.4$, $z>22.6$ and $y = 22.0 \pm 0.4$.

\section{Discussion}

\subsection{Redshift estimation}

\begin{figure}
    \centering
    \includegraphics[width=1\linewidth]{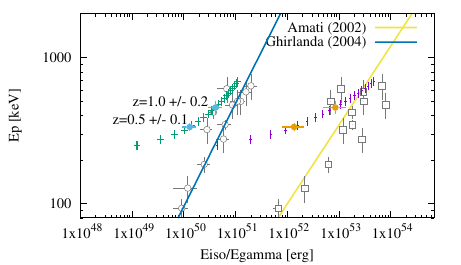}
    \caption{Position of the GRB in the Amati and Ghirlanda relations. Small points show the GRB positions for redshift varying from $z=0.1$ (leftmost) to $z=2.0$ (rightmost). Larger points highlight our two most likely redshift ranges: $0.38 < z < 0.56$ and $0.8 < z < 1.2$.}
    \label{fig:cos}
\end{figure}

With lack of direct spectroscopic measurement of GRB 200131A we are left to photometric estimates based on either optical afterglow photometry or host galaxy observations. In short, it can be said that from the positive detection by Swift-UVOT UVW2 and UVM2 filters that the redshift is relatively low, in range of $z\leq1.2$. If significantly larger than 1.2, the detection in the UVW2 filter would get much fainter comparing to the UVM2 filter, and as seen in Fig.\,\ref{fig:sed}, this is not the case. This argumentation is valid to $z\simeq0.8$, when the Lyman break gets out of the UVW2 filter. 
The somewhat fainter UVW2 can be also interpreted as Lyman-$\alpha$ reaching UVW2 and not UVM2, implying $0.38 < z < 0.56$. 

The host galaxy detection as we have it, 
scales to an absolute magnitude of the Milky Way (-21) at redshift $z=0.82$, providing a soft upper limit, as GRBs are usually not found in very large galaxies. 
The host is however far too weak to provide any further photometric redshift estimation. 

The observations by Konus-Wind \citep{gcnkonus} provide a peak energy of the
GRB's spectrum $E_\mathrm{p} = 228_{-15}^{+17}$ keV and fluence $f =
1.60_{-0.08}^{+0.09}\times10^{-5}$erg/cm$^2$. 
With Amati \citet{amati} relation we get
an estimation of redshift $z\simeq1.6$, but the spread of the relation is very permissive and, in fact, disfavors only redshifts smaller than $\sim$0.4.
%
%
With the assumed detection of the jet break in the lightcurve, we can further use the reversed Ghirlanda \citet{ghirlanda,campana07} relation to try and infer more redshift information. 
Ghirlanda relation is tighter than Amati, and can provide a somewhat stricter determination of redshift. 
We used the same approximation as the author, and similarly to Amati relation, obtain a solution that disfavours lower redshift bursts (see Fig.\,\ref{fig:cos}). 
%


Summing up, there seem to be two best windows for the redshift:
A. $ 0.38 < z < 0.56$ so that Ly-$\alpha$ is in UVW2 and not UVM2,
i.e. $z = 0.47 \pm 0.09$.
With this $z$, the host galaxy is moderately bright.
The burst is at an edge of Amati relation spread and gets too far from Ghirlanda relation.
B. $0.8 < z < 1.2$ which needs a minimum photometric contribution of Ly-$\alpha$ line to the UVOT UVM2 filter, and the defficiency in UVW2 is due to Lyman break at 912$\AA$. 
In the Amati and Ghirlanda relation, the burst fits well. 
The host galaxy is somewhat too bright in the upper limits of this interval. 

Taking into account all factors, our preferred redshift is at the lower end of the higher range ($z\sim0.9$).
This preference is based on several considerations:
At z > 1.2, the host galaxy would be unusually luminous for a GRB host
The lower range (0.38 < z < 0.56) would place this event among the nearest known long GRBs, which is statistically unlikely given the GRB redshift distribution.
The multi-wavelength properties of both prompt and afterglow emission are typical for long GRBs, suggesting the burst likely falls within the more common redshift range.
The observed spectral and temporal properties are consistent with those typically seen in GRBs at moderate redshifts.
While we use $z = 0.9$ in our modeling, we emphasize that this is a best estimate based on the available indirect evidence, and direct spectroscopic measurement would be required for a definitive determination.
The conclusions of our analysis regarding the reverse shock interpretation remain robust within the plausible redshift range of $0.8 < z < 1.2$.

\subsection{Afterglow fitting}

We started modelling of the afterglow with a simple power-law decay model, into which we added breaks as necessary. 
It turns out that two early breaks are necessary to fit the afterglow as well as a late break that we attribute to a jet break. 
Furthermore, the first and third segment of this simple fit are compatible with reverse and forward shock closure relations for a GRB afterglow.
We therefore abandoned this simplistic approach and rather take it as a supportive argument for validity of a physics based model.

We fitted the available optical (see Table\,\ref{tab:phot}) and X-ray \cite{ba}
afterglow behaviour in both time and frequency domain to a relativistic fireball
model \citep{piran}, particularly using relations presented by \citet{gao}. 
The model involves a forward shock with a hydrodynamic peak and a jet break superimposed at early times with a reverse shock with a common source in the expanding shell. 
The model includes the GRB redshift, but even if permitted to vary the event redshift, it cannot provide any useful estimate of $z$, so we fixed the redshift to $z=0.9$. 
The fit resulting parameters are summarised in Table\,\ref{tab:fit1}. 
The afterglow behaviour is consistent with a free expansion into a homogeneous interstellar matter (ISM) with the electron energy distribution parameter $p=2.3$ (see Figure~\ref{fig:lc}). 
The hydrodynamic peak parameters show significant uncertainties, with the forward shock temporal parameters being poorly constrained ($T_\mathrm{fs} = 33 \pm 22$,s, $G_\mathrm{fs} = 1.01 \pm 0.20$) and the rising power law index had to be fixed to 3.0 due to complete lack of constraint from the data. This is reflected also in the relatively large uncertainty of the forward shock magnitude scaling ($M_\mathrm{fs} = 14.00 \pm 0.33$,mag).
UVOT UVW2 ultraviolet points are influenced by Lyman forest blanketing, which is not implemented in the fit, so the UVW2 filter was left out of the fit. 
%
%
Also, any X-ray points from the first orbit of observations do not follow the model and had to be excluded from our fit. 

We found an indication for a late achromatic break in both X-ray and optical data. 
The available data points fit well with an achromatic break with properties expected from a jet break in an afterglow expanding into the ISM, i.e. the difference between decay slope before and after this break is $\Delta\alpha = 3/4$. 

Three groups of observations had to be assigned and fitted an independent zeropoint: All R-band observations from GCN require shifting by $B_\mathrm{R}=0.62$\,mag, the i' points by RATIR ($B_\mathrm{i'} = 0.19$\,mag) and, to our surprise, the photometric measurements by TLS Schmidt camera, which turn out 0.28\,mag too faint with respect to the GRB model. 
We note that the photometry published by \citet{gcnvirt} is off by precisely 2\,mag which we account to a typographic error in the GCN circular and use the point with a corrected value. 

During the afterglow fitting, we faced a problem with zeropoint incompatibility, $R$-band measurements seem to be universally fainter than $r'$, as well as $i'$ seem to be somewhat fainter than expected from the model. 
The wide-band TLS Gaia-like filter also shows $\sim 30\%$ fainter detection than expected. 
Some of these (R-band) may be linked to GCN data calibration problems, while the other may have origin in possible spectral features in the data.
We note, however, that our simple model does not take into account filter profiles.
Unless we have all the raw observational data in our hands, it is difficult to speculate of what could have caused these offsets. 

While most of the data follows well the standard afterglow model, the early X-ray data deviate from it and are systematically brighter than a prediction of our fitting. 
This may be accounted for a prolonged internal engine activity.
Also, as has been shown before in case of GRB 120326A \citep{urata}, as this X-ray  activity is somewhat correlated with the reverse shock in the optical band, it may have been produced by Synchrotron Self Compton radiation related to the contemporaneous reverse shock emission \citep{invcomp}. 
With the limited data, it is, though, difficult to say more.

\section{Conclusions}

Our analysis of GRB 200131A reveals a clear signature of reverse shock emission in the early optical afterglow, captured through rapid-response observations beginning just 63 seconds post-trigger. The afterglow evolution is well-described by a combination of reverse and forward shock components, with evidence for a late-time jet break, supporting the standard fireball model interpretation.

We observed an optical afterglow of GRB 200131A obtaining the first photometric point 63\,s after the satellite trigger. 
This early observation shows a steep decay, suggesting either internal engine activity or a reverse shock.
We complemented our observations with a critical selection of GCN-published photometric points. 
By fitting this data set, we show that the early data fit well as a reverse shock component of the GRB afterglow modelled as a thin shell expanding into a constant density interstellar matter. 
The fitting also shows a good agreement with a catalogued Milky Way galactic extinction and leaves only little space for further extinction in the host galaxy, as shown in Figure~\ref{fig:sed}. 

Although a direct measurement of the redshift for this gamma-ray burst (GRB) has not been obtained, there are several hints of GRB redshift. 
After judging UVOT detections, host galaxy detection, Amati \cite{amati} and Ghirlanda \cite{ghirlanda} relations for the GRB prompt and afterglow emission, we conclude that this GRB is likely to have occured at $0.8 < z < 1.0$. 
Only direct redshift measurement can provide definitive answer here and enable further interpretation of the data set. 
From the Pan-STARRS \cite{panstarrs} archival data it seems that the galaxy may be relatively easily accessible for spectroscopy with a large telescope. 

\bibliographystyle{actapoly-astro}
\bibliography{grb200131a} 

\end{document}